\documentclass{article}
\usepackage{spconf, amsmath, graphicx, enumitem, wrapfig, amssymb, xcolor, url, tabularx}

\setlist[itemize]{leftmargin=*, nosep}
\setlist[enumerate]{itemsep=1pt}

\title{DeepA: a deep neural analyzer for speech and singing vocoding}
\name{Sergey Nikonorov$^1$, Berrak Sisman$^2$, Mingyang Zhang$^1$, Haizhou Li$^{1,3}$}
\address{
  $^1$National University of Singapore, Singapore, 
  $^2$Singapore University of Technology and Design, Singapore \\
  $^3$The Chinese University of Hong Kong (Shenzhen), China}

\begin{document}
%
\maketitle


\begin{abstract}
Conventional vocoders are commonly used as analysis tools to provide interpretable features for downstream tasks such as speech synthesis and voice conversion. They are built under certain assumptions about the signals following signal processing principle, therefore, not easily generalizable to different audio, for example, from speech to singing. In this paper, we propose a deep neural analyzer, denoted as DeepA -- a neural vocoder that extracts F0 and timbre/aperiodicity encoding from the input speech that emulate those defined in conventional vocoders. Therefore, the resulting parameters are more interpretable than other latent neural representations. At the same time, as the deep neural analyzer is learnable, it is expected to be more accurate for signal reconstruction and manipulation, and generalizable from speech to singing. The proposed neural analyzer is built based on a variational auto-encoder (VAE) architecture. We show that DeepA improves F0 estimation over the conventional vocoder (WORLD). To our best knowledge, this is the first study dedicated to the development of a neural framework for extracting learnable vocoder-like parameters.
\end{abstract}



\section{Introduction}
\label{sec:intro}

Nowadays deep neural networks are used extensively in the fields of speech synthesis \cite{TTS_DNN, DeepVoice}, voice conversion \cite{VC, VC_SPC}, singing voice synthesis \cite{SVS, DAR} and so on. The neural solutions continue to use conventional vocoders, such as STRAIGHT \cite{STRAIGHT} and WORLD \cite{WORLD}, as they allow one to transform speech to and from a set of parameters, which can be manipulated independently. One common example of vocoder usage within a neural TTS framework is to train the network that maps input text into frames of vocoder parameters. For the training phase vocoder provides features that were proven to be an effective speech representation and after the model is trained the vocoder is effectively used as a synthesizer.

We note that conventional vocoders are designed under some assumptions about the properties of speech signals. A vocoder typically performs two functions -- signal analysis and reconstruction.  Recent studies show that neural reconstruction techniques, such as WaveNet \cite{WaveNet} and WaveRNN \cite{WaveRNN}, have surpassed conventional vocoders in terms of speech quality. This can be attributed to the data-driven techniques that allow the reconstruction function to learn from actual data distributions. 

The same strategy for neural reconstruction can naturally be applied for neural analysis. While there are some studies in related topics \cite{F0_NPM, LDRoTP, F0_COND, DDSP}, there is no study on a complete neural vocoder, that transforms input speech into a set of independent parameters which can then be used for speech reconstruction in a single neural framework. We have good reasons to believe that data-driven neural vocoder has its advantage over conventional vocoder, which will be the focus of this paper.

While the quality of the analysis part of the vocoder is sufficient for most of the tasks in the field of speech processing,  there are tasks in the field that conventional vocoders struggle with. One such task is singing voice synthesis 
where the reconstructed singing quality dropped significantly when F0 is changed by more than an octave scale.  Investigating this problem carefully led us to the following conclusion: vocoder parameters do not allow good quality speech reconstruction for the case of significant change in F0. There turned out to be three main possible causes for that:
\begin{enumerate}
    \item inconsistency of the vocoder parameters
    \item errors in F0 estimation
    \item the assumption that timbre is independent from pitch
\end{enumerate}
The third point (cause 3) above is a result of the fact that conventional vocoders were designed as speech encoding and processing tools -- in case of low F0 modulation the assumption holds. Further experiments implied that in order to account for the timbre dependency on pitch linguistic information of the processed speech sample is required. Causes 1 and 2, however, arise from the errors in vocoder algorithm. They should be leveraged by learnable features and a sufficient amount of appropriate data, which motivates the studies in this paper. In this paper, we make a step towards a neural analyzer suitable for singing vocoding by addressing points 1 and 2 from the list above.

The main contribution of this paper is the development a neural speech analyzer, i.e., DeepA, that generates consistent and concise features. DeepA allows for a high-quality singing voice reconstruction and outperforms the conventional vocoder in terms of F0 estimation. To our best knowledge, this is the first study dedicated to neural analyzer-synthesizer pipeline.

This paper is organized as follows: section II is dedicated to the related work, section III provides a more detailed analysis of the issues mentioned above, section IV describes the proposed framework, in section V experimental results and their interpretation are presented. Finally section VII concludes the study.


\section{Related work}
Before an in-depth problem analysis, it is useful to review both conventional and neural solutions to speech analysis and reconstruction to motivate our proposal. Conventional vocoders were originally developed as speech processing tools for voice compression and transformation (hence the name ``vocoder", which stands for ``voice encoder"). They represent input signals with vocoder parameters, and reconstruct output signals from parameters. The parameterization allows for manipulation of the physical properties of input signals. The two functions are referred to as analysis and synthesis.

\subsection{Conventional vocoders: analysis \& synthesis}
The early vocoders did not achieve good quality of speech reconstruction, however that changed with the development of STRAIGHT \cite{STRAIGHT} and then WORLD \cite{WORLD}. They achieve this by decomposing speech into three temporal sequences of parameters -- the fundamental frequency (F0), smoothed spectrogram (roughly corresponds to the voice timbre) and aperiodicity (correlates with ``voice quality" attribute \cite{VSV_ext}). This strategy allowed for high-quality speech reconstruction and (within a certain range) manipulation.

Conventional vocoders are still used today, mainly either as voice manipulation tools (for example, for changing F0 in voice conversion tasks) or as the provider of interpretable features for training various neural frameworks \cite{VAE_VC, NU-NAIST, hn-NSF}. However, being speech processing tools built upon certain assumptions about the analyzed signal, conventional vocoders are not flexible enough to be used for high-quality singing voice synthesis. Even though both WORLD and STRAIGHT were either extended to be able to solve the task \cite{VSV_ext} or applied directly (with some additional tools) \cite{UTAU}, both approaches still use a lot of assumptions and heuristics. It should be mentioned that parameter estimation from data via numerical optimization is often employed by those systems, but the number of these parameters is usually quite small, which makes the whole system to rely on the highly constrained model. While a well-engineered model in tandem with a relatively restricted numerical parameter estimation can outperform the more general data-driven neural frameworks in a task the system was designed for, the latter system is more flexible in terms of application and is more adaptable to new data.

In this paper we focus on utilizing these properties of neural frameworks in combination with structural constraints, inspired by conventional vocoders, to learn a compact singing voice representation, which allows for both the accurate speech reconstruction and manipulation.

\subsection{Neural vocoders: synthesis}
Recent development in the field of neural networks showed that those systems are capable of high-quality speech synthesis, WaveNet \cite{WaveNet}, WaveGlow \cite{WaveGlow}, WaveRNN \cite{WaveRNN}, GAN-based models \cite{MelGAN, VocGAN}, LPCNet \cite{LPCNet}, neural homomorphic vocoder \cite{NHV} and NSF models \cite{hn-NSF} being the few examples. Since they reconstruct speech using F0 and/or other acoustic features (e.g. spectrogram, phoneme encoding etc.) a term ``neural vocoder" was coined to describe such systems. The architecture and idea behind each neural vocoder may differ (for example, WaveNet generates raw audio, sample by sample, in autoregressive manner, while hn-NSF essentially implements a harmonic-plus-noise model \cite{HNM} within a neural framework), but all of those systems are flexible in terms of input features.

However, they are not ``vocoders" in  full sense of the word as they do not perform speech analysis, but rather rely on another module or algorithm, such as STRAIGHT, WORLD, and short-time Fourier transform, for feature extraction. In that regard such frameworks are essentially neural synthesizers, which rely on the assumption that extracted features are adequate and are free from errors. Even though some of the frameworks attempt to mitigate errors in the input features, speech manipulation using such systems remains a challenge. For example, hn-NSF model is able to deal with F0 errors to some extent, but when combining the modified F0 and the original spectrogram as an input, the resulting speech sample has a pitch that is different from what can be inferred from either of the input parameters.

Given this issue and the fact that there appears to be no study dedicated to the development of neural vocoder in full sense, i.e., a complete neural analyzer and synthesizer, the focus of this paper is the development of such a neural vocoder system. In particular, we employ hn-NSF model as a neural synthesizer to be trained jointly with a neural analyzer, architecture of which was inspired by both conventional vocoders and the hn-NSF model itself.

\subsection{Variational auto-encoder: representation learning}

Auto-encoder \cite{AE} is a system that aims to learn the latent, usually -- more compact and/or meaningful, representation of the input. Auto-encoders may be viewed as systems that perform dimensionality reduction, but in contrast to principal component analysis the components of the encoding can be constrained or ``conditioned" during the training, making the network to converge to the representation with desired properties. For example, the encoding can be conditioned to be ``smooth" in a way that makes sense in terms of the task at hand, which may be used for data denoising \cite{IR}.

Variational auto-encoder (VAE) \cite{VAE} extends the basic architecture with an additional constraint -- the obtained latent representation should resemble a sample from a certain well-known distribution. Gaussian distribution is a common choice for the distribution in question, since an expression for Kullback–Leibler divergence (which is used as part of the loss) between two Gaussian distributions is quite simple and can be derived analytically. The main idea of VAE is to implicitly estimate the distribution of the input data and learn the effective encoding based on that information.

VAE-based models are widely used for learning disentangled representation of some parameters, that may be used for reconstructing the input. For example, it was used for learning the disentangled representation of pitch and timbre of musical instruments \cite{LDRoTP} and for learning the latent representation of speaking style in a TTS framework \cite{StyleControl}. The framework proposed in this paper is built as a variational auto-encoder in order to learn independent representations of harmonic and noise parts of the input sample, which allows one to use the system for speech decomposition, manipulation and reconstruction, similar to the conventional vocoders.



\begin{figure}[t!]
    \centering
    \includegraphics[width=\columnwidth]{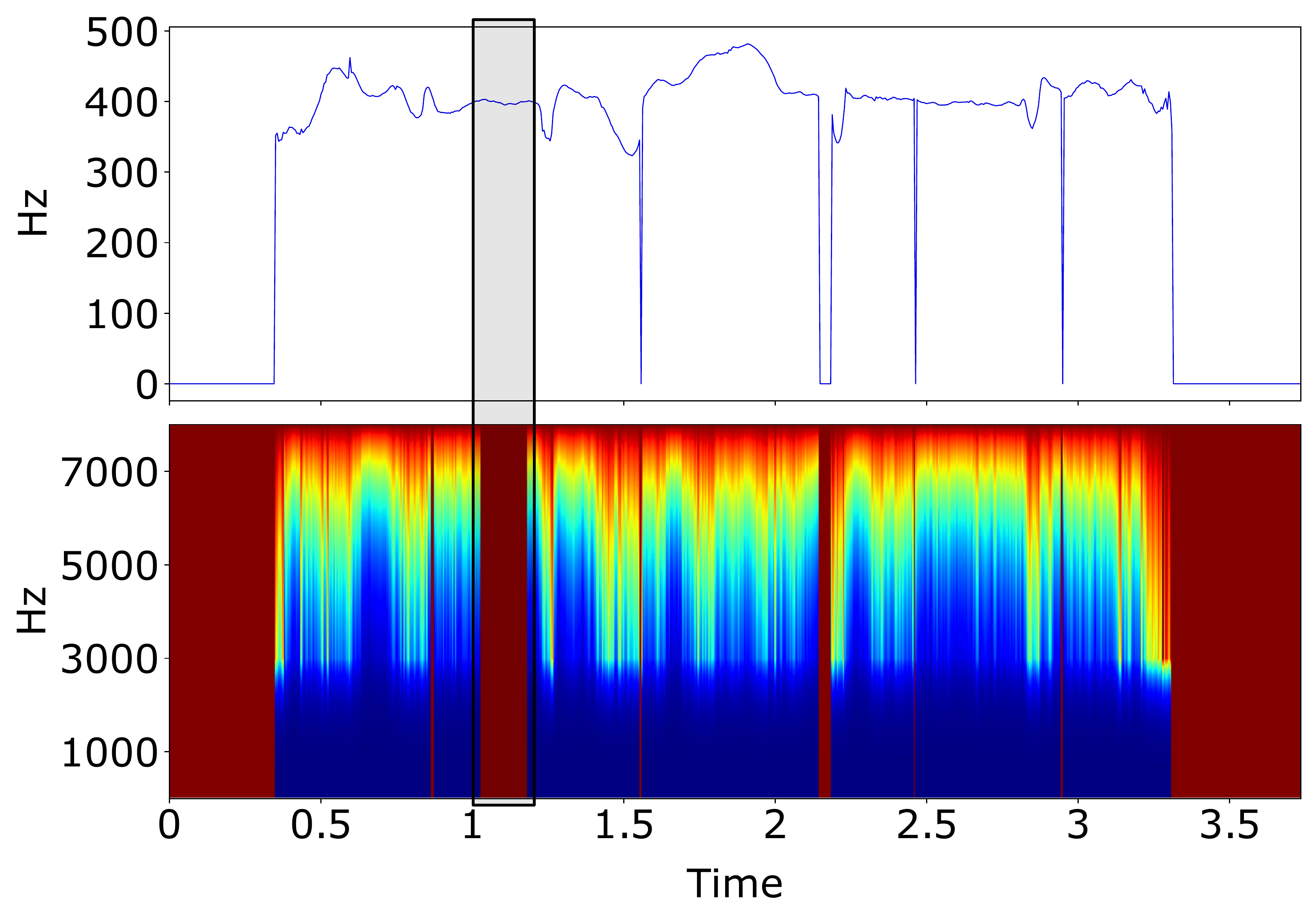}
    \caption{F0 (top) and aperiodicity (bottom) of the same sample derived by WORLD vocoder. Notice that according to F0 the highlighted frames are voiced while according to aperiodicity they are not.}
\end{figure}

\section{Issues with conventional vocoders}

We would like to look into two issues with conventional vocoder algorithms, which adversely affect the quality of singing voice manipulation (causes 1 and 2 mentioned in the introduction). The analysis is focused on the WORLD vocoder \cite{WORLD}, which is used extensively and shares common characteristics with STRAIGHT \cite{STRAIGHT}~\cite{Rev_SER}.

\subsection{Inconsistency of vocoder parameters}

Vocoder parameters are designed to be independent, but they are expected to be consistent with each other. For example, both F0 and aperiodicity generated by the WORLD vocoder have clearly defined voiced and unvoiced segments and ideally those segments should overlap each other exactly. However, sometimes that is not the case. As shown in Fig. 1, F0 and aperiodicity, calculated by WORLD \cite{WORLD}, may predict different ``voicedness" (V/UV flag) of the same frame range (highlighted by a grey rectangle in the picture). The sample, reconstructed using these parameters, exhibits distortion in the interval of inconsistency.

\begin{figure}[t]
    \centering
    \includegraphics[width=\columnwidth]{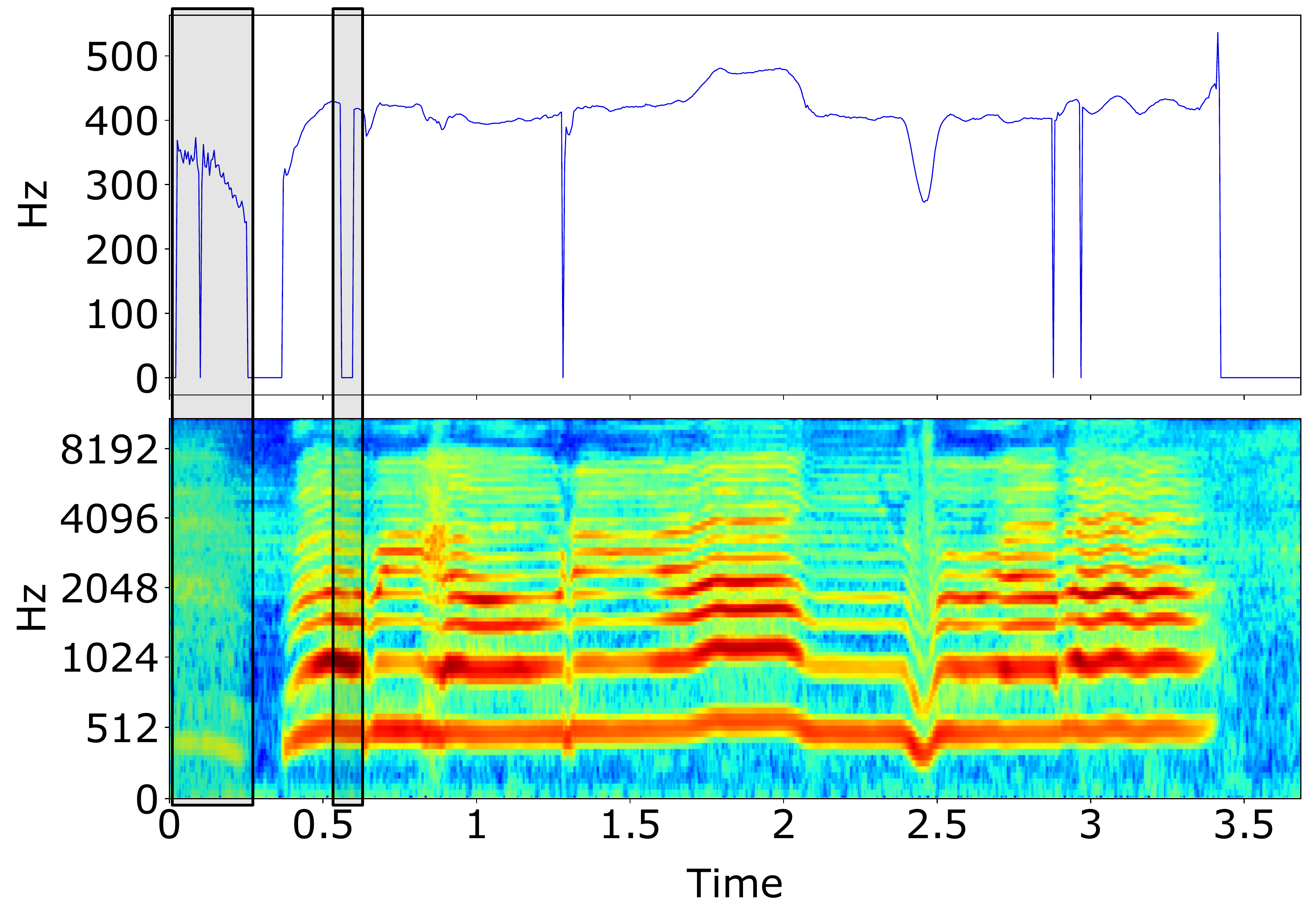}
    \caption{F0 (top) and mel-spectrogram (bottom) of the same sample derived by WORLD vocoder. Notice that the implicitly estimated V/UV flag of the highlighted frame ranges doesn't tally with the spectrogram. }
\end{figure}

\subsection{Errors in F0 estimation}

Even when F0 is consistent with aperiodicity, the former may still contain noticeable errors. As shown in Fig. 2, the first group of highlighted frames (at the very beginning of the sample) has no periodic structure according to the mel-spectrogram. However, the vocoder estimated F0 of that range as close to the mean pitch of the sample. The second highlighted group exhibits the opposite problem -- according to F0 the frames are unvoiced even though the spectrogram clearly shows the continuous periodic structure in that area.

\begin{figure*}[t!]
    \centering
    \includegraphics[width=\textwidth]{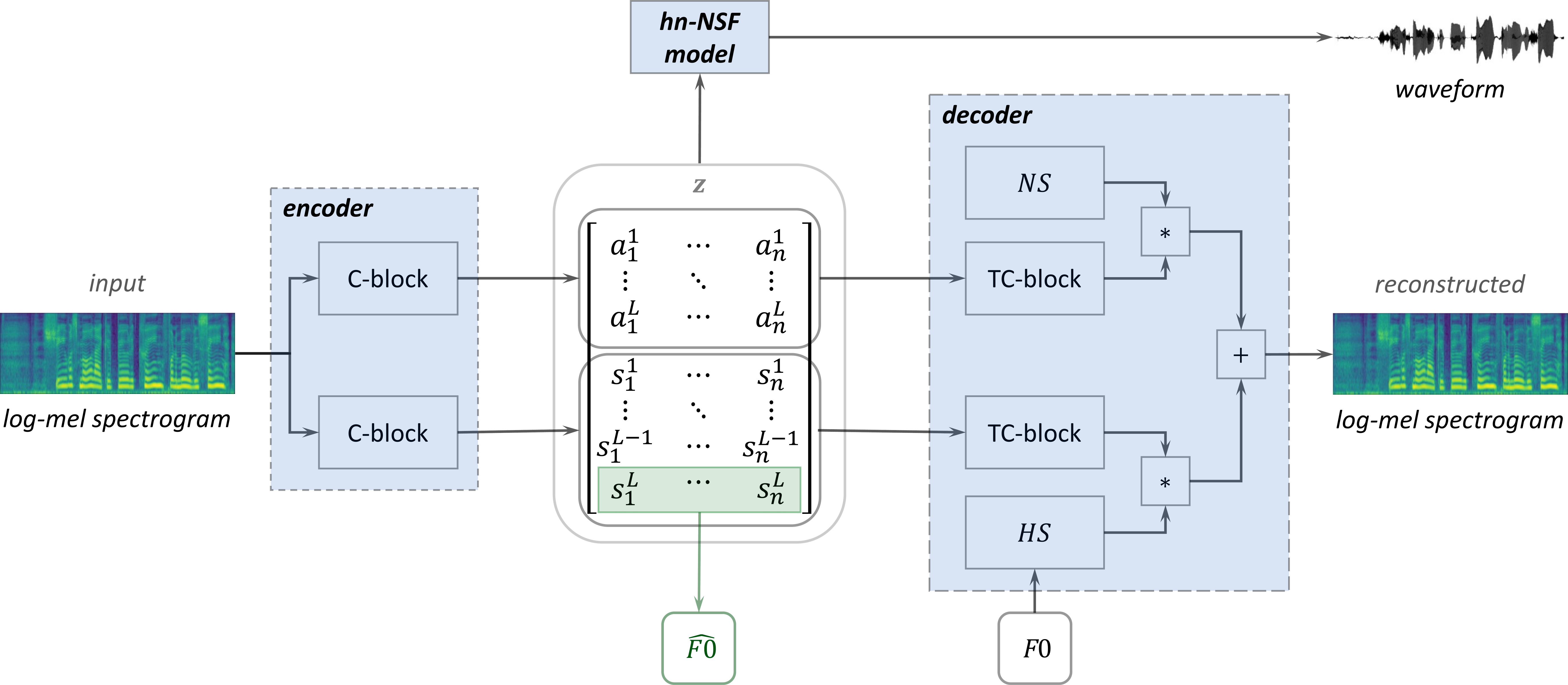}
    \caption{The proposed neural analyzer, DeepA, that is based on a VAE architecture.}
\end{figure*}

A vocoder seeks to decompose speech or singing vocals into interpretable vocoder parameters. While text-to-speech systems may allow rare mild distortions, the high dynamic range of the singing voice makes those distortions more prominent, which may significantly degrade the perceived singing quality. Such distortions in conventional vocoders usually arise from hard-coded human assumptions in the system.  We believe that a neural analyzer with data-driven modeling training could be a solution, that we will discuss next. 


\vspace{-\baselineskip}
\section{Deep Neural Analyzer}

The proposed neural analyzer, denoted as DeepA for ``Deep Analyzer", is trained jointly with a neural synthesizer. This allows the neural analyzer and synthesizer to interact during training: the synthesizer can guide the analyzer with its speech reconstruction loss, while the analyzer conditions the synthesizer to effectively utilize the latent representation via the modified VAE and spectrogram reconstruction loss.

DeepA follows an  encoder-decoder VAE structure, which is coupled with a hn-NSF \cite{hn-NSF} synthesizer (see Fig. 3) during training. After the system is trained on speech data, the encoder and hn-NSF model can be used as a speech analyzer and synthesizer respectively. The hn-NSF model is chosen as a synthesizer here because it allows for explicit control of F0 when generating speech. With hn-NSF synthesizer, as long as the F0 parameter is consistent with the rest of the input parameters, the F0 of the generated speech is guaranteed to match the F0 control signal. This is especially important for singing, where a fine fluctuation of pitch in the singing voice \cite{F0_dynamics} may greatly affect the overall voice quality.  


\begin{figure}[th]
    \centering
    \includegraphics[width=\columnwidth]{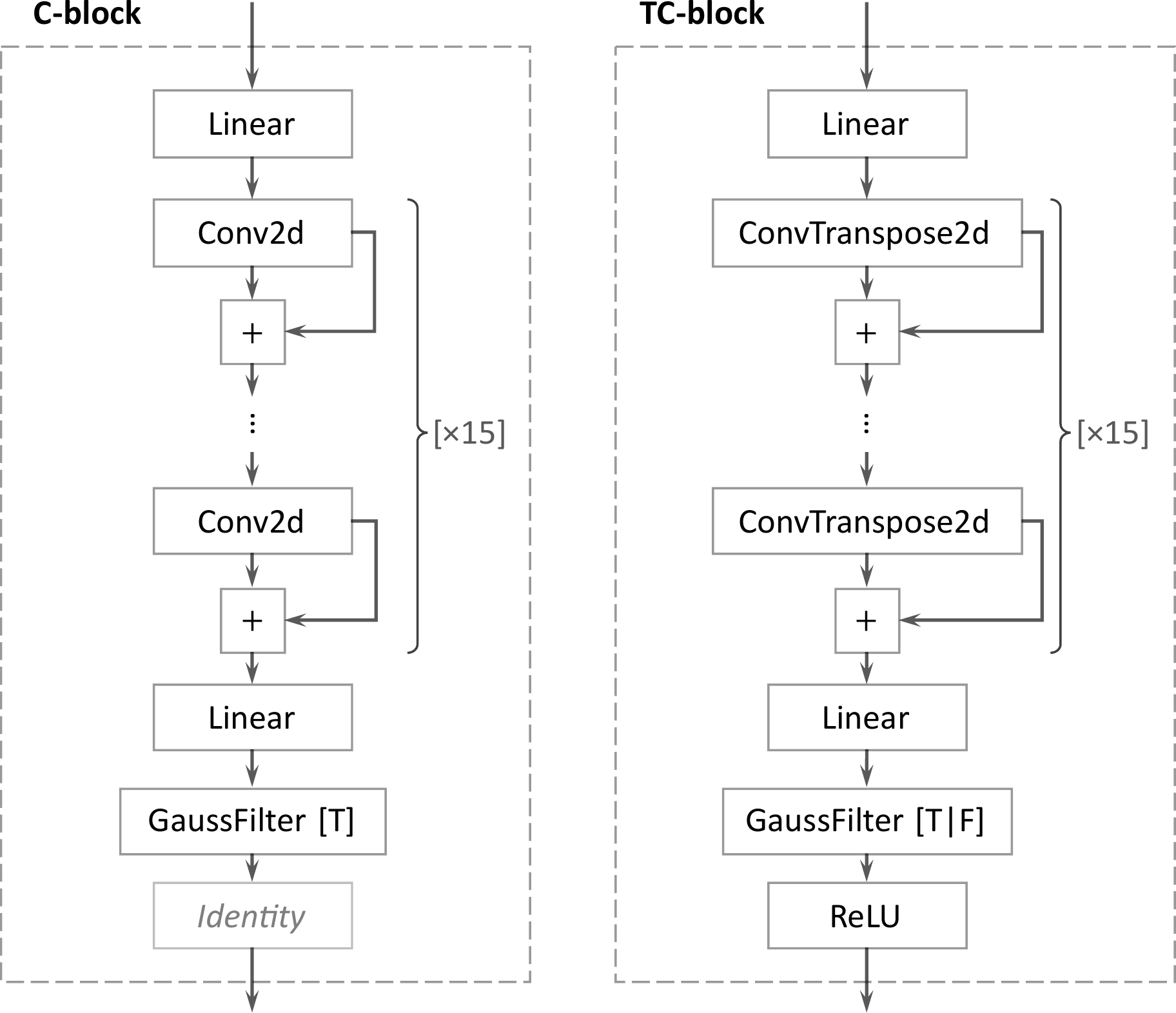}
    \caption{C-block (left) and TC-block (right) structure}
\end{figure}

\subsection{Encoder}
The encoder consists of two parallel C-blocks (see Fig. 4), which map the input mel-spectrogram to the latent representations of the noise (matrix $a$ in a diagram) and harmonic (matrix $s$ in a diagram) parts of the input. Both matrices are $L$ rows by $T$ columns, where $L$ is the latent dimension and $T$ is the number of temporal frames in the input. The full latent representation $z \in \mathbb{R}^{2L \times T}$ is defined as a vertical stack of $a$ and $s$. It should also be mentioned that the last row of the matrix $s$ (which conveniently corresponds to the last row of the latent representation itself) is conditioned to resemble the normalized ground truth F0. The encoder architecture is inspired by a hn-NSF \cite{hn-NSF} neural synthesizer, which uses two separate blocks for generating harmonic and noise parts of the speech signal independently and then adds them together to produce the final output. The same idea is used for the decoder architecture as well, which is described in the next subsection.

\subsection{Decoder}
As can be seen from the diagram (Fig. 3), the decoder is structurally similar to encoder, using TC-blocks (see Fig. 4) instead of C-blocks. The latent speech representation $z$, which is passed to the decoder as an input, is decomposed into matrices $a$ and $s$, which are then processed by the decoder separately. This structural constraint is proven to be essential for learning disentangled representations of harmonic and noise parts of the spectrogram. The decoder transforms those representations into masks, which are then multiplied by the generated white noise mel-spectrogram (\textit{NS} block in Fig. 3) or a mel-spectrogram of a harmonic signal with ground truth F0 as a fundamental (\textit{HS} block in Fig.3).


\subsection{Training details}

We are particularly interested in the performance of F0 estimation and rendering. Therefore, we conduct experiments on singing voice samples from the NUS-HLT Speak-Sing (NHSS) database \cite{NHSS}, which constitutes approximately 7 hours of audio. All samples were downsampled to 16 kHz, mel-spectrograms of order 80 were used as input features, ground truth F0 was extracted using the WORLD \cite{WORLD} vocoder and the latent dimension $L$ was experimentally chosen to be 16. The framework also uses the modified VAE loss in order to improve the system stability. The full loss of the system is given by the following expression: 
\begin{equation}
    \begin{split}
        Loss = \alpha_1\cdot&\|F0 - \widehat{F0}\|^2 + \alpha_2\cdot Loss_{KL}(\mu, \nu) + \\
               + &\|x - \hat{x}\|^2 + Loss_{NSF}(y, \hat{y})
    \end{split}
\end{equation}
where $Loss_{KL}$ is the averaged over time Kullback-Leibler divergence between $\mathcal{N}(\mu, e^\nu)$ and $\mathcal{N}(0, I)$, which is essentially the distance between two distributions (see \cite{VAE} for derivation)
and $Loss_{NSF}$ is a hn-NSF synthesizer loss defined in \cite{hn-NSF}. Basically, it is a sum of three STFT losses between the ground truth waveform and the generated one, which represents the speech reconstruction loss in frequency domain. The first quadratic term of $Loss$ represents F0 reconstruction loss, and the third quadratic term of $Loss$ represents the VAE reconstruction loss (mean squared error between the input mel-spectrogram and the reconstructed one).
Here $T$ is the number of temporal frames in the input, $\mu$ and $\nu$ denote the predicted mean and log-variance of the latent representation $z$; $F0, x$ and $y$ denote ground truth F0, mel-spectrogram and singing waveform respectively, while the corresponding circumflexed variables ($\widehat{F0}, \hat{x}$ and $\hat{y}$) denote the generated ones. 
$\alpha_1$ and $\alpha_2$ are positive weights determined experimentally. The discovered working configuration is $(\alpha_1 = 10^2, \alpha_2 = 10^{-2})$. 


\vspace{-0.5\baselineskip}
\section{Experiments}

We perform both objective and subjective evaluation. The objective evaluation was performed on a subset of 195 utterances from NHSS dataset \cite{NHSS} (not seen by the system during training), reported  in terms of  mel-cepstral distortion (MCD) and F0 deviation, computed in two different ways (RMSE and median of absolute difference). The subjective evaluation was performed in a form of a mean-opinion score (MOS) test based on the overall voice quality.

\vspace{-0.5\baselineskip}
\subsection{Experimental setup}

In this paper three systems are compared: WORLD vocoder, hn-NSF synthesizer using WORLD vocoder parameters, and the proposed neural analyzer followed by hn-NSF synthesizer. WORLD vocoder parameters were extracted using 5 ms frame period and 1024 frequency bins (for spectrogram and aperiodicity), a baseline hn-NSF synthesizer used F0 extracted by WORLD and mel-spectrogram of order 80 and DeepA used just a mel-spectrogram of order 80 with the hn-NSF synthesizer within the proposed framework using latent spectrogram representation $z$ of order 32 as its input.


\subsection{Objective evaluation}


\textbf{MCD:}
\par\noindent
Mel-cepstral distortion (MCD) is a perceptual-based measure of similarity between two speech samples \cite{MCD}. The MCD statistics for the three systems are summarized in Table 1. It is observed that all the tested models achieved comparable quality in terms of of speech reconstruction. As the MCD metric was shown to correlate with perceived speech quality \cite{MCD}, this result gives a rough idea of the actual quality of the proposed framework relative to the established models. 

\begin{table}[h]
    \setlength{\tabcolsep}{8pt}
    \begin{center}
        \begin{tabular}{|c|c|c|c|}
            \hline
            System & {Analyzer} & {Synthesizer} & {MCD} \\
            \hline
            \hline
            1 & WORLD & WORLD & 5.2 ± 1.0 \\
            \hline
            2 & WORLD & hn-NSF & 4.2 ± 0.6 \\
            \hline
            3  & DeepA & hn-NSF & 5.1 ± 0.5 \\
            \hline
        \end{tabular}
        \caption{MCD [dB] (mean ± std) value comparison among 3 systems.}
    \end{center}
\end{table}


\begin{figure*}[t]
    \centering
    \includegraphics[width=\textwidth]{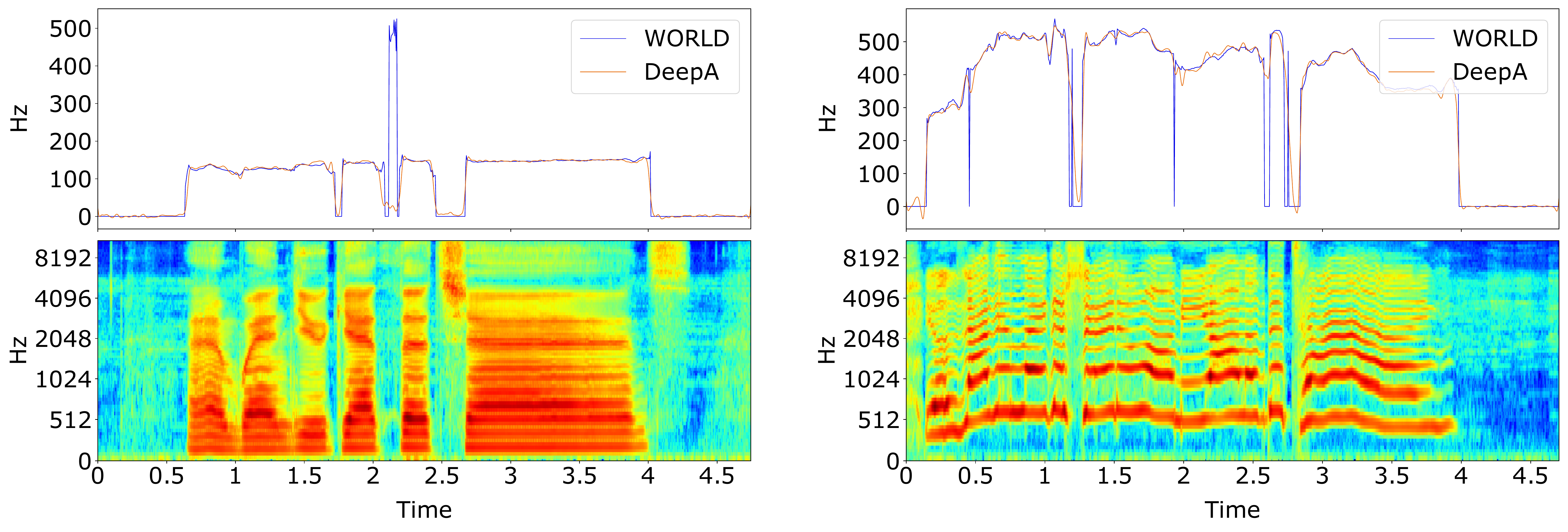}
    \caption{F0 (top row) and mel-spectrogram (bottom row) of two samples, sung by a male (left) and a female (right) singer, with F0 derived by WORLD and DeepA respectively.  In the F0 plots, we note that DeepA provides a more appropriate V/UV flag than WORLD, that can be verified against the mel-spectrogram. }
\end{figure*}

\vspace{-0.5\baselineskip}
\noindent\textbf{F0 deviation:}
\par\noindent
Two methods were used to calculate the deviation between F0 extracted by WORLD and F0 generated by the proposed model -- root mean square error (RMSE) and median of absolute difference (further referred to as MD). The expression for MD metric is given below:
\begin{equation}
    MD(F0, \widehat{F0}) = median(|F0 - \widehat{F0}|)
\end{equation}
where $F0$ and $\widehat{F0}$ are fundamental frequency contours extracted by WORLD and generated by the proposed model respectively. The results of F0 deviation are as follows:

\vspace{3pt}
\begin{raggedleft}
    \begin{tabular}{l l}
        \textbf{RMSE}: & 28.0 ± 13.0 Hz \\
        \textbf{MD}: & 4.0 ± 1.2 Hz
    \end{tabular}
\end{raggedleft}
\vspace{6pt}


As can be seen from F0 deviation values, the RMSE value appears to be quite large. This is due to the fact that RMSE is sensitive to outliers. When several samples with the highest RMSE values were investigated, it turned out that the large error is caused by those frames, for which the WORLD vocoder and DeepA predict the opposite V/UV flag. This made RMSE a  very useful indicator to highlight the discrepancy between the two systems. By inspecting 20 worst samples in terms of RMSE, it turned out that in almost all the cases the proposed model predicted the correct V/UV flag (two examples of the F0 contour comparison are shown in Fig. 5).

To accurately represent the overall performance of the proposed framework, the median of absolute difference (MD) was employed. MD is insensitive to outliers, thus giving the robust approximation of the central tendency. The MD value for the DeepA is 4 Hz, which, in conjunction with the observations mentioned in the paragraph above, shows that the system is both in good agreement with WORLD vocoder and robust, predicting more accurate results when WORLD fails.

\vspace{-0.3\baselineskip}
\subsection{Subjective evaluation}

We report the mean-opinion score (MOS) in Table 2. This test seeks to determine the overall quality of speech generated by WORLD, hn-NSF (which uses WORLD F0 and mel-spectrogram as input) and the proposed framework (DeepA analyzer + hn-NSF synthesizer) with respect to the ground truth. 10 subjects participated in the test and each listened to 96 utterances in total. From Table 2 it can be seen, that the proposed framework achieved the quality comparable to the other systems. These results together with the objective evaluation in the previous subsection suggest that the proposed framework is capable of both analyzing and reconstructing singing voice with appropriate quality. The advantage of the neural analyzer-synthesizer pipeline lies in the fact that it employs interpretable latent representation that allows direct  F0 manipulation in the same way as conventional vocoders. 
\begin{table}[h]
    \setlength{\tabcolsep}{8pt}
    \begin{center}
        \begin{tabular}{|c|c|c|c|}
            \hline
            System & {Analyzer} & {Synthesizer} & {MOS } \\
            \hline
            \hline
            \multicolumn{3}{|c|}{\textit{Ground truth}} & 4.1 ± 0.5 \\
            \hline
            1 & WORLD & WORLD & 3.8 ± 0.5 \\
            \hline
            2 & WORLD & hn-NSF & 3.9 ± 0.5 \\
            \hline
            3 & DeepA & hn-NSF & 3.7 ± 0.6 \\
            \hline
        \end{tabular}
        \caption{MOS value (mean ± std) on 96 utterances.}
    \end{center}
\end{table}


\vspace{-1.7\baselineskip}
\section{Conclusion}

We propose DeepA, a neural analyzer-synthesizer pipeline as a novel vocoder. It was able to achieve singing quality comparable with WORLD and hn-NSF model in both objective and subjective evaluations. We have seen that DeepA provides more accurate and robust F0 estimation than WORLD, that is required for singing vocoding. As the performance of neural analyzer is on par with that of WORLD, this work marks an important step towards an interpretable neural vocoder. 


\bibliographystyle{IEEEbib}
\bibliography{refs}

\end{document}